\documentclass[lettersize,journal]{IEEEtran}
\usepackage{amsmath,amsfonts}
\usepackage{algorithmic}
\usepackage{algorithm}
\usepackage{array}
\usepackage[caption=false,font=normalsize,labelfont=sf,textfont=sf]{subfig}
\usepackage{textcomp}
\usepackage{stfloats}
\usepackage{url}
\usepackage{verbatim}
\usepackage{graphicx}
\usepackage{cite}
\usepackage{amsmath,amsfonts}
\usepackage{algorithmic}
\usepackage{algorithm}
\usepackage{array}
\usepackage[caption=false,font=normalsize,labelfont=sf,textfont=sf]{subfig}
\usepackage{textcomp}
\usepackage{stfloats}
\usepackage{url}
\usepackage{verbatim}
\usepackage{graphicx}
\usepackage{cite}
\usepackage{amssymb}
\usepackage{pifont}
\usepackage{float}
\usepackage{tabulary}
\usepackage{makecell}
\usepackage{threeparttable}
\usepackage{booktabs}
\usepackage{makecell}
\begin{document}

\title{A Novel Geometry Method for LED Mapping}

\author{Junlin Huang, Shangsheng Wen and Weipeng Guan
\thanks{This work was supported  by  the
National Undergraduate Innovative and Entrepreneurial Training Program
under Grant 202110561162. {\it{ (Corresponding author: Shangsheng Wen.)}}}

\thanks{Junlin Huang, Shangsheng Wen and Weipeng Guan are with the School of Materials Science
and Engineering, South China University of Technology, Guangzhou 510641,
China (e-mail: shshwen@scut.edu.cn).}

}

\markboth{}%
{Shell \MakeLowercase{\textit{et al.}}: A Sample Article Using IEEEtran.cls for IEEE Journals}


\maketitle

\begin{abstract}
With inputs from RGB-D camera, industrial camera and wheel odometer, in this letter, we propose a geometry-based detecting method, by which the 3-D modulated LED  map can be acquired with the aid of visual odometry (VO) algorithm from ORB-SLAM2 system when the decoding result of LED-ID is inaccurate.  Subsequently, an enhanced cost function is proposed to optimize the mapping result of LEDs. The average 3-D mapping error (8.5cm) is evaluated with a  real-world experiment. This work can be viewed as a preliminary work of Visible Light Positioning (VLP) systems, offering a novel way to prevent the labor-intensive manual site surveys of LEDs.

\end{abstract}


\section{Introduction}
\IEEEPARstart{V}{isible} light positioning (VLP)  systems   have attracted a great research interest in recent decades \cite{ref1,ref2,ref3,ref4}, for its high transmission rate, pinpoint accuracy and no electromagnetic interference \cite{ref5,ref6,ref7} play a significant role in indoor positioning field. By utilizing light beam from light-emitting diode (LED), which carries unique identity code (LED-ID) and frequency, VLP systems can unscramble information related to LED location for further pose estimation of robot. To achieve this, most VLP systems \cite{ref8,ref9,ref10,ref11} require a pre-built map composed of global LED locations and  identifiers, for which manual site surveys are labor-intensive. There are a handful of works aiming to map LED locations. In \cite{ref12,ref13}, the authors propose to exploit a mobile robot equipped with a 2-D LiDAR and a rolling-shutter camera for VLP calibration. The robot approaches each LED, takes images of over head LED and decodes LED-ID. The LiDAR data processed by a Simultaneous Localization and Mapping (SLAM) algorithm give the 2-D LED position with an accuracy of centimeters, whereas the height of LED still need a mannual survey. Recently, a LED mapping system, LedMapper, has been proposed \cite{ref14}. The system is evaluated on a self-assembled  handheld mapping device with visual-inertial sensors, including two rolling-shutter cameras and an inertial measurement unit. Extensive experiments verify its efficacy and performance in building 3-D LED map by posing a full-SLAM problem within a factor graph formulation. But the system still requires a few LEDs surveyed manually as control points.

Another difficulty which affects the performance of VLP systems is that the fragile transmission channel due to unpredictable environmental variation and the immanently asynchronized air conveying channel will impede the decoding of LED-ID \cite{ref15}. Plenty of thresholding methods, e.g., quick adaptive threshold, polynomial threshold and iterative threshold, have been proposed to overcome this dilemmas, whereas fail in adapting different transmission channel \cite{ref16}.   

Instead of forcing an absolutely precise and stable  decoding of LED-ID, in this letter, we propose a geometry detecting method based on inputs from RGB-D camera, industrial camera and wheel odometer, by which the 3-D modulated  LED map can be acquired with the aid of visual odometry (VO) algorithm from ORB-SLAM2 system \cite{ref17}, when the decoding result of LED-ID is inaccurate.  The only usage of LED-IDs is  classifying different LEDs for building LED map. By solving a least square problem with a proposed enhanced function, the detected position of  LED  can be optimized and  the mapping accuracy (8.5cm error on average) is evaluated by an indoor experiment. As we use modulated LEDs, this work should be viewed as a pre-work for VLP systems.


\section{Principle}

A robot equipped with RGB-D camera (Cam1), industrial camera (Cam2) and wheel odometer will execute the LED mapping  task.  In the following discussion, we use superscript $\cdot^{obs}$, $\cdot^{pos}$ to denote the parameter belonging to observation and posterior function respectively, and
subscript $\cdot_{k}$ is exploited to  denote the parameter belonging to $k_{th}$ observation. We highlight that all the LEDs are modulated and  LED-IDs should be different but no correctly decoded LED-ID is necessary. The only use of LED-IDs is to distinguish between different LEDs.

\begin{figure}[!t]
	\centering
	\includegraphics[width=3.1in]{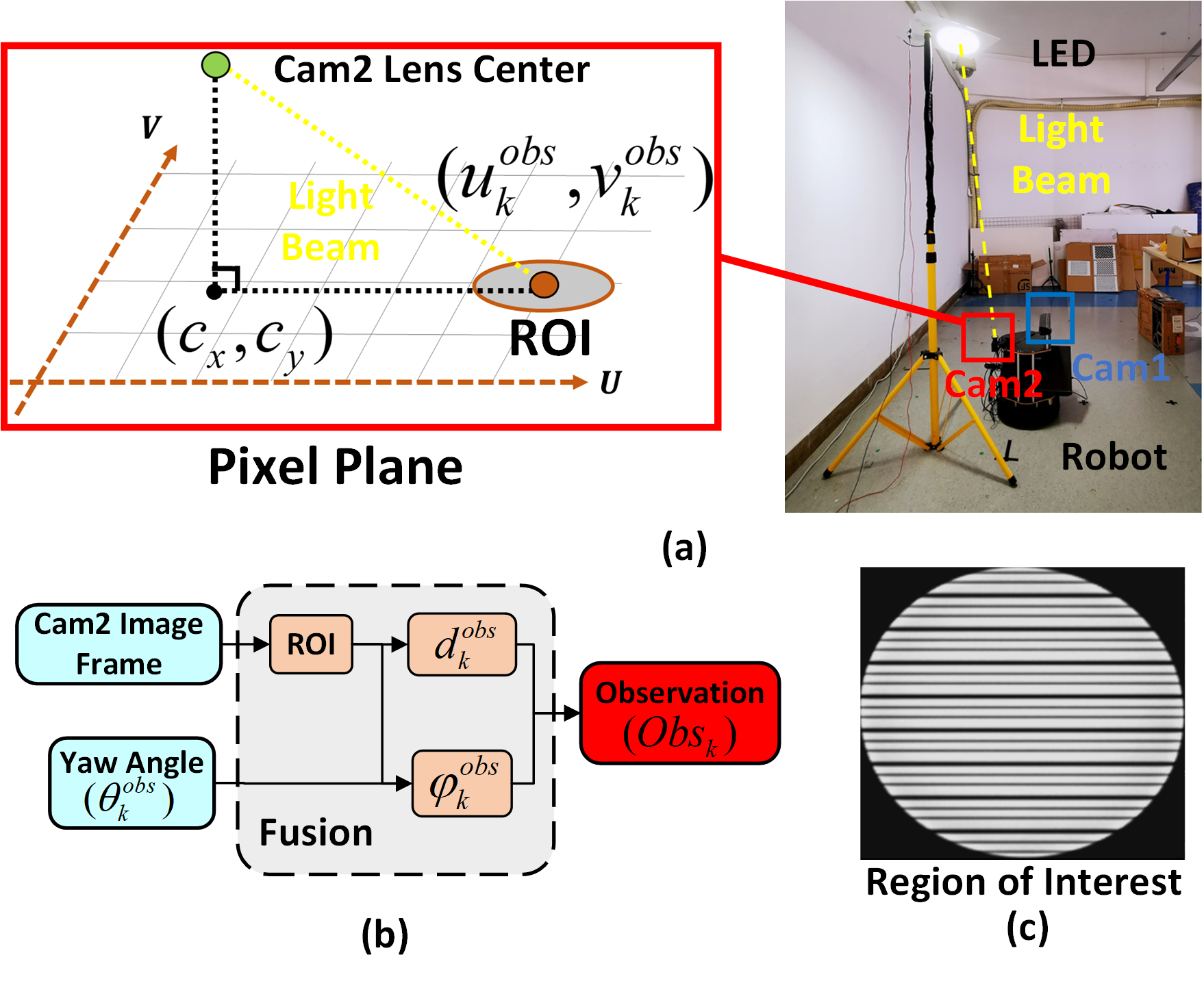}
	\caption{(a) Geometrical relationship in industrial camera. (b) A  data fusion scheme. (c) A snapshot of Region of Interest.}
	\label{2D1}
\end{figure}

\begin{figure}[!t]
	\centering
	\includegraphics[width=3.6in]{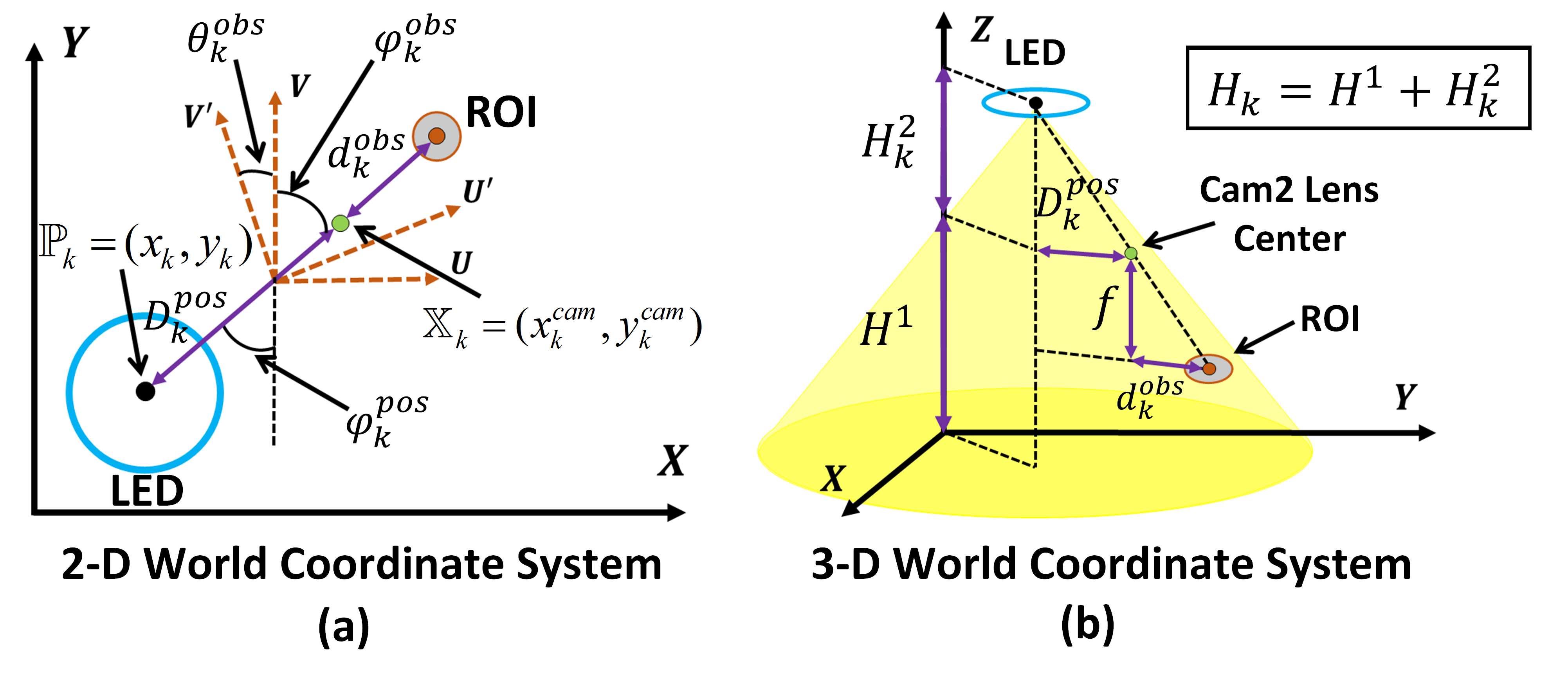}
	\caption{(a) The parameters defined in posterior function and their geometrical relationship with observed parameters. (b) The elevation relationship between LED  and Cam2. }
	\label{2}
\end{figure}

A  data fusion scheme is proposed (Fig. \ref{2D1}(b)).   From Cam2 image frame, the Region of Interest (ROI) and  $d^{obs}_{k}$, which describes 2-D distance between Cam2 lens center and ROI center,  can be obtained.  
Coupled with yaw angle $\theta^{obs}_{k}$ from wheel odometer,  the observation is defined as
\begin{equation}
\begin{split}
Obs_{k}&=\begin{pmatrix} 
\sqrt[2]{(u^{obs}_{k}-c_x)^2+(v^{obs}_{k}-c_y)^2}\\
arctan(u^{obs}_{k}-c_x,v^{obs}_{k}-c_y)-\theta^{obs}_{k}\\
\end{pmatrix}\\
&=\begin{pmatrix}
d^{obs}_{k}\\
\varphi^{obs}_{k}\\
\end{pmatrix}\\
\end{split}
\end{equation}
where $(c_x,c_y)$ represents the coordinate of Cam2 lens center projected on pixel plane, which is an intrinsic parameter, and $(u^{obs},v^{obs})$ is the ROI center coordinate on pixel plane. Fig. \ref{2D1}(a) and Fig. \ref{2}(a) illustrate the parameters defined in $Obs_{k}$. 

Subsequently, a thresholding method is applied to LED  position detection. The 2-D coordinate of Cam2 $\mathbb{X}_{k}$ and 
2-D coordinate of LED  $\mathbb{P}_{k}$ are defined as
\begin{equation}
\begin{split}
\mathbb{X}_{k}&\triangleq(x^{cam}_{k},y^{cam}_{k})\\
\mathbb{P}_{k}&\triangleq(x_{k},y_{k})\\
\end{split}.
\end{equation}
Because the observed position of Cam1 can be obtained from VO algorithm of ORB-SLAM2 directly, the 2-D pre-estimated position of Cam2 denoted by
\begin{equation}
\begin{split}
\tilde{\mathbb{X}}_{k}&\triangleq(\tilde{x}^{cam}_{k},\tilde{y}^{cam}_{k})
\end{split}
\end{equation}
can be acquired by a fixed coordinate transformation between Cam1 and Cam2 (green fixed transformation in Fig. \ref{exp2}(f)).
Next, the Cam2 position from $k_{th}$ observation $\tilde{\mathbb{X}}_{k}=(\tilde{x}^{cam}_{k},\tilde{y}^{cam}_{k})$ is considered to be an alternative estimated position of LED  $\mathbb{P}^{alt}_{k}=(x_{k}^{alt},y_{k}^{alt})$ when $d_{k}^{obs}$ is smaller than an artificial threshold, i.e.,
 $\mathbb{P}_{k}^{alt}=\tilde{\mathbb{X}}_{k}$. By applying Pauta criterion\footnote{A well-known data processing principle, also referred to as $3\sigma$ principle.}  to all the previously observed alternative estimated positions of LED, we can acquire the roughly estimated position of LED $\tilde{\mathbb{P}}_{k}=(\tilde{x}_{k},\tilde{y}_{k})$.

Next, a geometry method is proposed in order to acquire a posterior function. Fig. \ref{2}(a) describes the parameters used in posterior function. The 2-D distance from LED  center to  Cam2 lens center is recorded as  $D^{pos}_{k}$ and the vertical angle of $\varphi^{obs}_{k}$ is recorded as $\varphi^{pos}_{k}$. Next, by similar triangle rule,  we introduce a parameter $\mathbf{k}_{k}$ from the elevation relationship between LED  and Cam2 in  Fig. \ref{2}(b)
\begin{equation}
\begin{split}
\mathbf{k}_{k}
&\triangleq\frac{d_{k}^{obs}}{D_{k}^{pos}}\\ 
&=\frac{\sqrt[2]{(u_{k}^{obs}-c_x)^2+(v_{k}^{obs}-c_y)^2}}{\sqrt[2]{(\tilde{x}_{k}^{cam}-\tilde{x}_{k})^2+(\tilde{y}_{k}^{cam}-\tilde{y}_{k})^2}}.\\
\end{split}
\end{equation}
Similarly,  by applying the Pauta criterion to all the previously observed  $\mathbf{k}_{k}$'s,  we can  acquire a refined parameter $\tilde{\mathbf{k}}_{k}$. Notice that $f$, the focal length of Cam2,  is an intrinsic parameter and $H^1$, the height of Cam2, can be measured in advance, we can calculate the height of  LED  $H_{k}$ in $k_{th}$ observation with the following equation
\begin{equation}
H_{k}=H^{1}+H^{2}_k=H^{1}+\frac{f}{\tilde{\mathbf{k}}_{k}}.
\end{equation}
Then, a posterior function $Pos_{k}(\mathbb{X}_{k},\mathbb{P}_{k})$ is designed to rectify the $k_{th}$ observation $Obs_{k}$

\begin{equation}
\begin{split}
Pos_{k}(\mathbb{X}_{k},\mathbb{P}_{k})&=\begin{pmatrix} 
\tilde{\mathbf{k}}_{k}*\sqrt[2]{(x_{k}^{cam}-x_{k})^2+(y_{k}^{cam}-y_{k})^2}\\
arctan(x_{k}^{cam}-x_{k},y_{k}^{cam}-y_{k})\\
\end{pmatrix}\\
&=\begin{pmatrix}
\tilde{\mathbf{k}}_{k}*D_{k}^{pos}\\
\varphi_{k}^{pos}\\
\end{pmatrix}
\end{split}
\end{equation}
where $\mathbb{X}_{k}=(x_{k}^{cam},y_{k}^{cam})$, $\mathbb{P}_{k}=(x_{k},y_{k})$ are unknown variables.

Then, an enhanced cost function consisting of $Obs_{k}$ and $Pos_{k}(\mathbb{X}_{k},\mathbb{P}_{k})$, which will contribute to the optimization of  LED  position, is proposed.
%
An error function is defined as follows
\begin{equation}
\begin{split}
	&error(\mathbb{X}_{k},\mathbb{P}_{k})\\
\triangleq&||Obs_{k}-Pos_{k}(\mathbb{X}_{k},\mathbb{P}_{k})||_2\\
=&\sqrt[2]{errdist(\mathbb{X}_{k},\mathbb{P}_{k})^2+errang(\mathbb{X}_{k},\mathbb{P}_{k})^2}
\end{split}
\end{equation}
where
\begin{equation}
\begin{split}
errdist(\mathbb{X}_{k},\mathbb{P}_{k})
\triangleq&d_{k}^{obs}-\tilde{\mathbf{k}}_{k}*D_{k}^{pos}\\
=&\sqrt[2]{(u_{k}^{obs}-c_x)^2+(v_{k}^{obs}-c_y)^2}-\\
&\tilde{\mathbf{k}}_{k}*\sqrt[2]{(x_{k}^{cam}-x_{k})^2+(y_{k}^{cam}-y_{k})^2}\\
\end{split}
\end{equation}
and
\begin{equation}
\begin{split}
	errang(\mathbb{X}_{k},\mathbb{P}_{k})\triangleq&\varphi_{k}^{obs}-\varphi_{k}^{pos}\\
=&(arctan(u_{k}^{obs}-c_x,v_{k}^{obs}-c_y)-\theta_{k}^{obs})\\
	&-arctan(x_{k}^{cam}-x_{k},y_{k}^{cam}-y_{k}).
\end{split}
\end{equation}
We substitute $\tilde{\mathbb{X}}_{k}$ for $\mathbb{X}_{k}$  in (7) and propose an enhanced cost function
\begin{equation}
J_1(\mathbb{P}_{k})=\sum_{l=1}^k\frac{1}{2}||error(\tilde{\mathbb{X}}_{l},\mathbb{P}_{k})||^2.
\end{equation}
By solving the following least squares problem
\begin{equation}
	\hat{\mathbb{P}}_{k} \in \arg{\min}J_1(\mathbb{P}_{k})
\end{equation}
we can acquire the optimized position of LED  $\hat{\mathbb{P}}_{k} =(\hat{x}_{k},\hat{y}_{k})$. Together with $H_k$ in (5), we acquire the 3-D position of LED  in $k_{th}$ observation $(\hat{x}_{k},\hat{y}_{k}, H_k)$, which can be used in VLP systems for robot localization.

\begin{figure}[!t]
	\centering
	\includegraphics[width=3.4in]{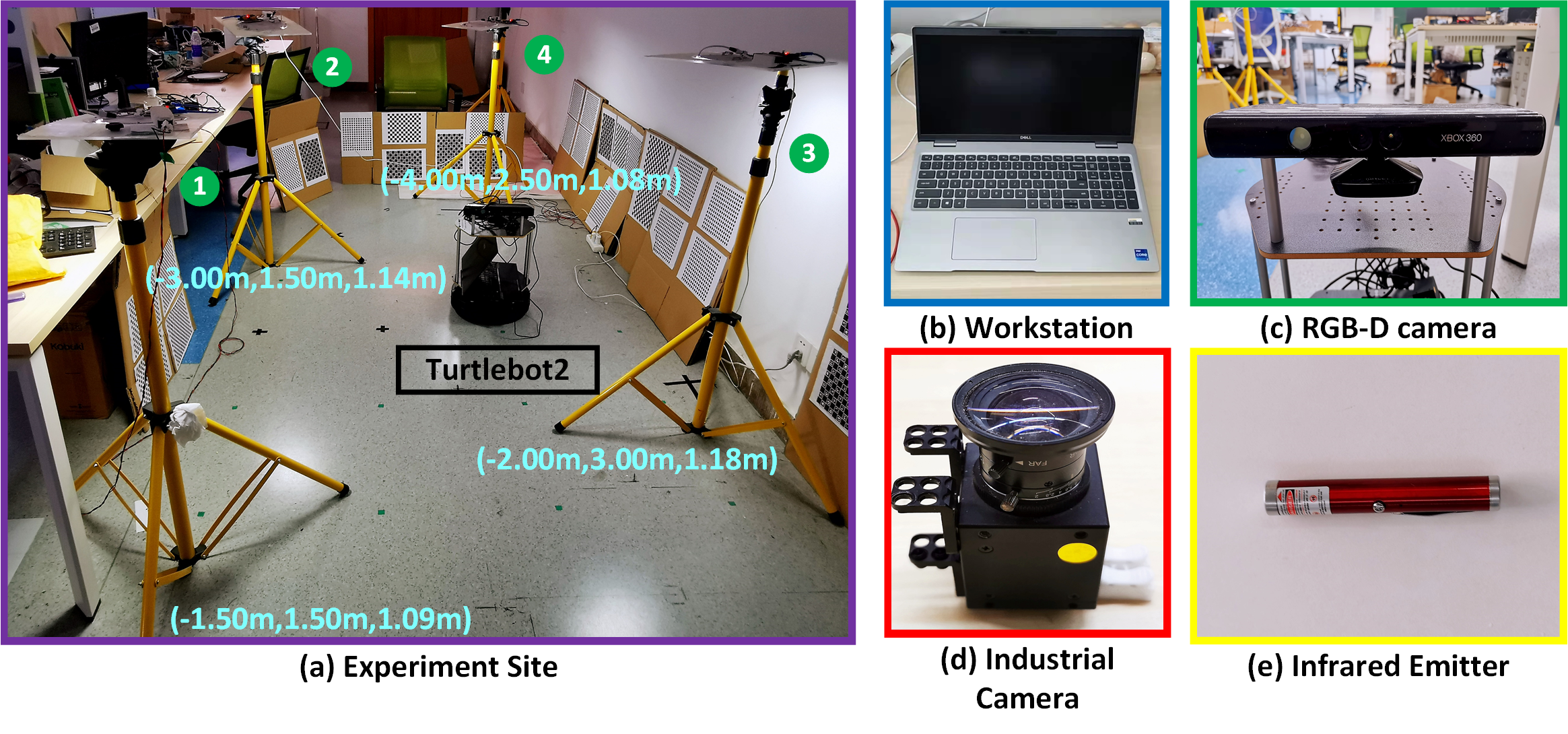}
	\caption{Hardware setup.}
	\label{3}
\end{figure}

\section{Experiment and Discussion}
In this module, an indoor  experiment is carried out in order to verify the mapping accuracy of our proposed method together with the LED  map construction performance.

As shown in Fig. \ref{3}, a two-wheeled differential driving mobile robot, Turtlebot2, equipped with a Kinect V1 For Xbox360 RGB-D camera , a MindVision UB-300 industrial camera  with prior extrinsic calibration and a laptop Module HASSE CW85S07 with Intel(R) Core(TM) i5-9400 and 8.00 GB RAM  is the main instrument to execute LED  mapping task. A remote control workstation  Module DELL Precision 3561 with 11th Gen Intel(R) Core(TM) i7-11850H and 32 GB RAM sends the operational command to robot with the aid of Secure SHell connection on Ubuntu 16.04  LTS.

\begin{figure}[!t]
	\centering
	\includegraphics[width=3.3in]{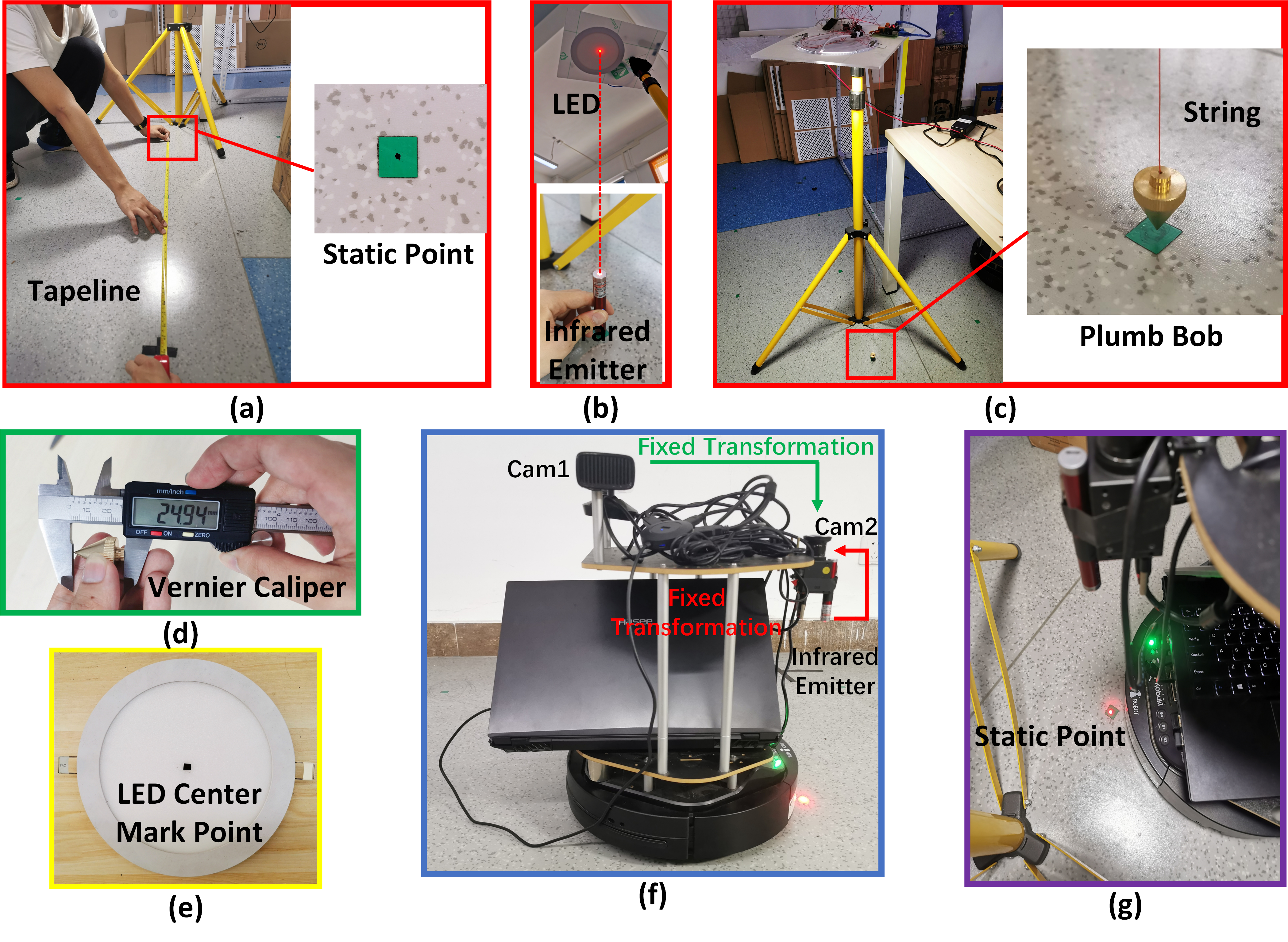}
	\caption{ LED position measurement process.}
	\label{exp2}
\end{figure}

Four LEDs of the same specification are used in experiment (Fig. \ref{3}(a)). In order to acquire the exact positions of the four LEDs, we measure and mark four static points on the world coordinate system (Fig. \ref{exp2}(a)). We then mark the LED center and install it on anchor (Fig. \ref{exp2}(e)).    
Next, an infrared emitter is placed on the static point and we let the infrared ray shines on the mark point of LED, which means that the projection point of LED center on the horizontal plane and the static point coincide (Fig. \ref{exp2}(b)). We hang a plumb from the LED so that the end of plumb just touches the ground (Fig. \ref{exp2}(c)).  Then we remove the plumb, measure the diameter of the plumb  (Fig. \ref{exp2}(d)) and the length of the string, and add the two numbers together to get the height of LED. The above measurements are made with tapeline  and vernier caliper. We treat the manual survey result as ground truth. Subsequently, we combine infrared transmitter with industrial camera and shoot it vertically on the ground to form a red dot. There is a fixed coordinate transformation between the infrared emitter and the lens center of the industrial camera  (red fixed transformation in Fig. \ref{exp2}(f)). Through this fixed coordinate transformation, the coordinate of the red point can be converted to the coordinate of the lens center of the industrial camera. Therefore, in order to achieve the positioning of LED, we only need to control the robot to move the red point to the static point  (Fig. \ref{exp2}(g)). We get the measurement error by subtracting the ground-truth value from the measured value and the error accuracy is 0.1cm because the minimum scale of tapeline is 0.1cm.

\begin{figure}[!t]
	\centering
	\includegraphics[width=3.5in]{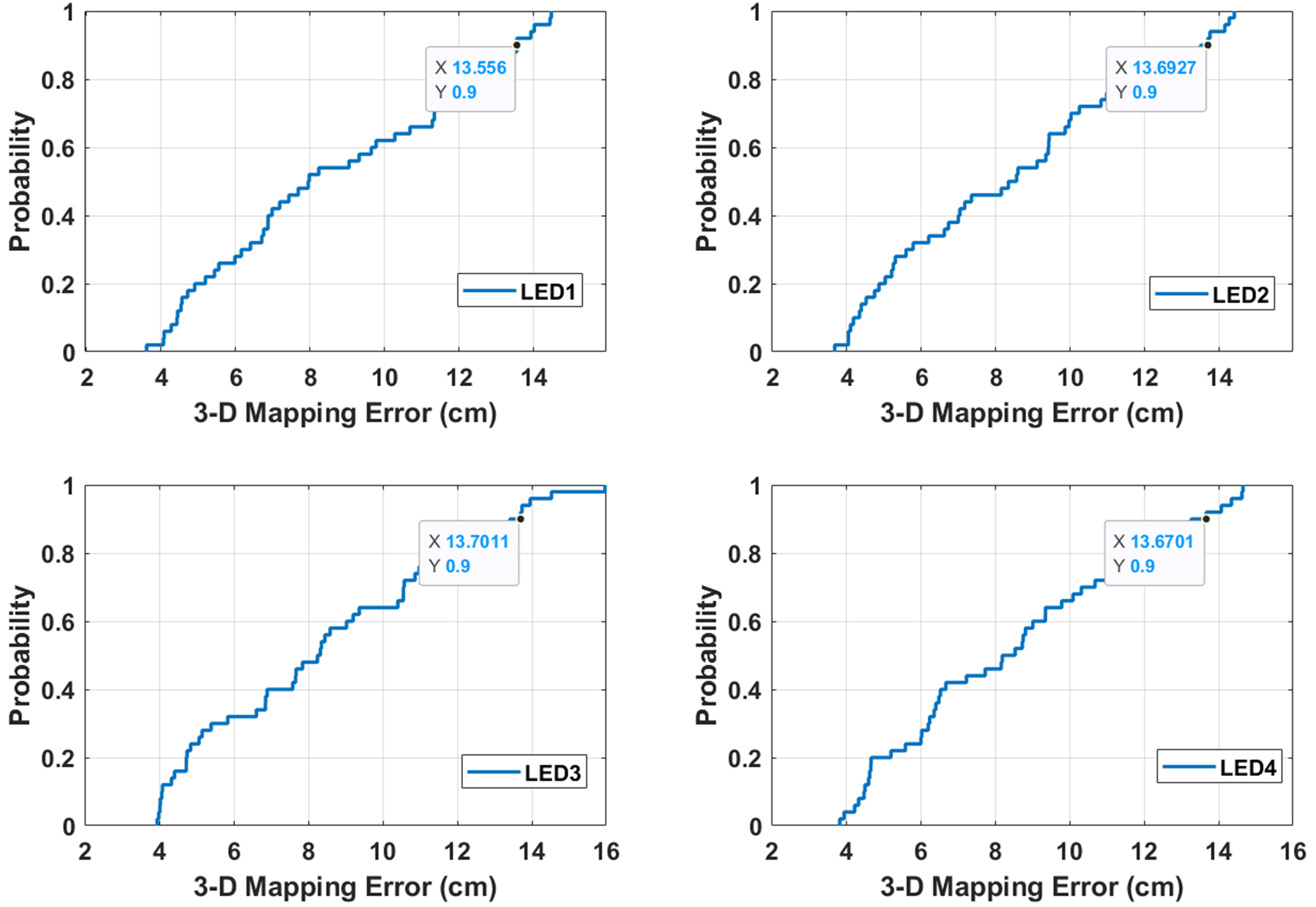}
	\caption{ Cumulative distribution function (CDF) curves of 3-D LED mapping error.}
	\label{4}
\end{figure}

\begin{figure}[!t]
	\centering
	\includegraphics[width=3.0in]{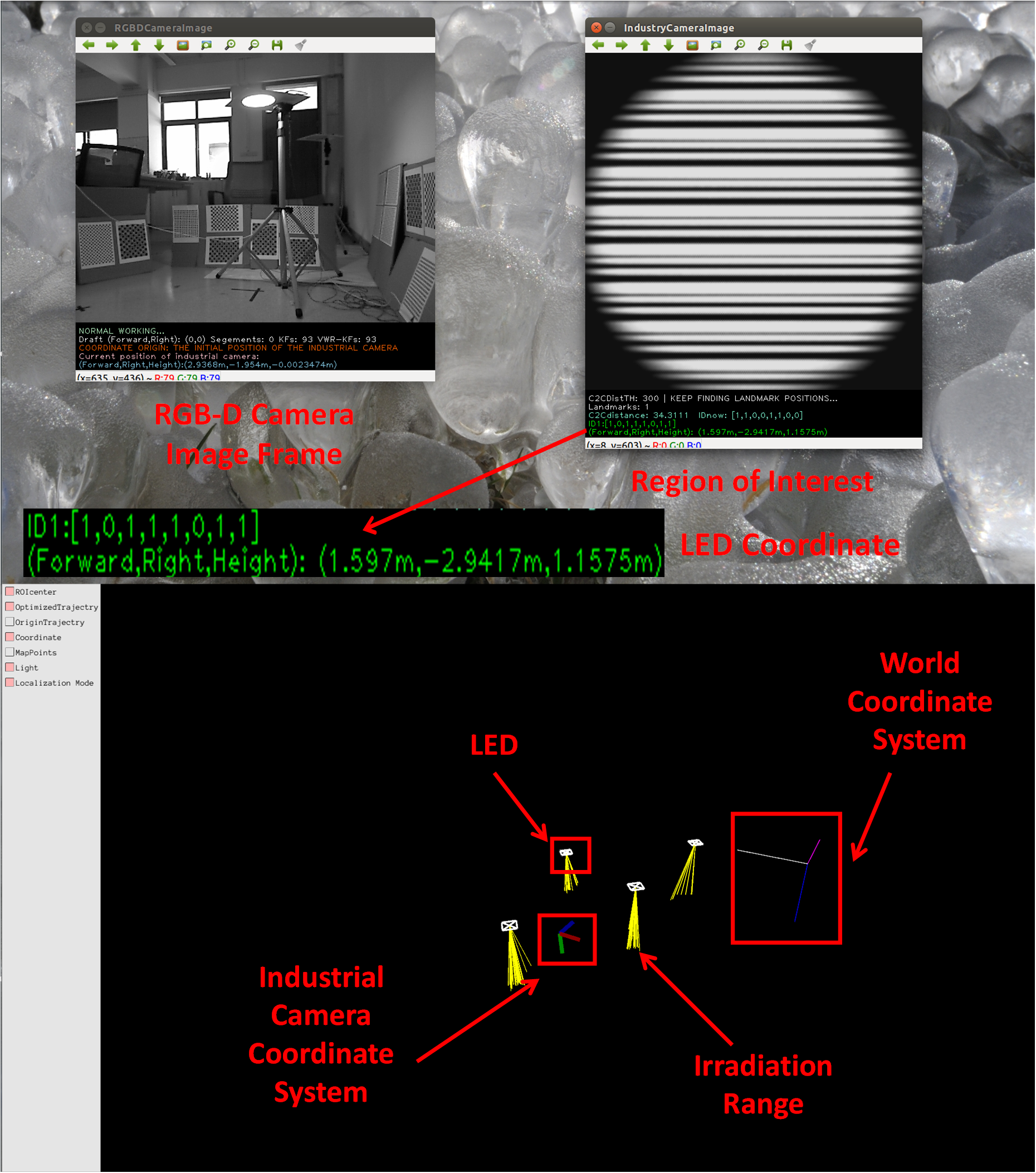}
	\caption{ A snapshot of LED map construction result.}
	\label{map}
\end{figure}

 In the experiment,  we artificially set the departure position of robot to be the origin of world coordinate system so that RGB-D camera position can be acquired by VO algorithm from ORB-SLAM2. Each LED's position is measured 50 times.   As shown in Fig. \ref{4}, more than 90\% 3-D mapping errors are less than 13.8cm. The confidence interval (90\% confidence level) of the 3-D mapping error is $[8.1,8.9]$cm, and the average  error is 8.5cm. Moreover, the LED  map is built and visualized with the aid of Pangolin\footnote{A lightweight library belonging to OpenGL for graphics drawing.}  library. Fig \ref{map} illustrates a snapshot of map construction result.

 We compare the performance of the proposed geometry method with the state-of-the-art (SOTA) works in LED mapping field. The average mapping error, correctly decoded LED-ID and receiver type are displayed objectively in Table \ref{tb2}. Compared with works in \cite{ref12,ref13}, our method can achieve 3-D LED mapping accuracy with 8.5cm average error. It is obvious that there is no need for our method to force a correctly decoded LED-ID, which means that our method is more stable and robust when faces some unpredictable fragile transmission environments. Besides, the  odometer price is much  lower than that of LiDAR and IMU, which indicates a low-cost deployment in wide adoption.

\section{Conclusion}
In this letter, we propose a  LED map construction method for VLP systems based on geometrical relationship between LED and industrial camera. By utilizing inputs from RGB-D camera, industrial camera and wheel odometer, the location of LED can be detected without a guarantee of precisely decoded LED-ID. In order to optimize the acquired LED position, an enhanced cost function consisting of observation and a posterior function is proposed, turning optimization step into solving a least square problem. Moreover,  an indoor experiment is conducted so as to verify the  mapping accuracy (8.5cm error)  together with LED map construction performance of our method, and the comparison with SOTA works demonstrates its advantages.


\begin{table}[!t]
\begin{center}
    \caption{Performance of Proposed LED Mapping Method}
    \vspace{1pt}
    \centering
\begin{threeparttable}
    \begin{tabular}{ccccc}
\toprule[2pt]
        \thead[c]{Method} & \thead[c]{Average\\Mapping\\ Error}& \thead[c]{Correctly \\Decoded\\ LED-ID} & \thead[c]{Receiver Type}\\
        \midrule[2pt]

Ref.\cite{ref12,ref13} &6cm (2D)& required& Camera+LiDAR \\

Ref.\cite{ref14} &2.2cm (3D)&required & Camera+IMU\\ 

Ref.\cite{ref18} &$\geq$ 10cm (3D)&required & PD receiver\\

\bf{Our Method} &8.5cm (3D)& not required& Camera+odometer\\
\bottomrule[2pt]  
    \end{tabular}
    \label{tb2}
\begin{tablenotes}
\footnotesize
\item[*]IMU: Inertial Measurement Unit.
\item[*]PD: Photodiode.

 \end{tablenotes}
\end{threeparttable}
\end{center}

\end{table}

\end{document}